\documentclass[journal]{IEEEtran}

\usepackage{epstopdf}
\usepackage{mhchem}
\usepackage{upgreek}
\usepackage{hyperref}
\usepackage{breakurl}
\usepackage{graphicx}
\usepackage{enumerate}
\usepackage{supertabular}

\newcounter{Rownumber}

\begin{document}

\title{Design and Implementation of High-throughput PCIe with DMA Architecture between FPGA and PowerPC}

\author{Kun~Cheng,Weiyue~Liu, Qi~Shen and Shengkai~Liao,

\thanks{Kun Cheng was with the Department of Modern Physics and Hefei National Laboratory for Physical Sciences at Microscale, University of Science and Technology of China, Hefei 230026, China.}
\thanks{Weiyue Liu was with Ningbo University, Ningbo 315211, China.}
\thanks{Qi Shen and Shengkai Liao are with the Chinese Academy of Sciences (CAS) Center for Excellence and Synergetic Innovation Center in Quantum Information and Quantum Physics, University of Science and Technology of China, Shanghai 201315, China}
\thanks{Email: skliao@ustc.edu.cn}}

\markboth{Journal of IEEE Transactions on Nuclear Science}
{shell \MakeLowercase{\textit{et al.}}:Bare Demo of IEEEtran.cls for IEEE Journals}

\maketitle

\begin{abstract}
We designed and implemented a direct memory access (DMA) architecture of PCI-Express(PCIe) between Xilinx Field Program Gate Array(FPGA) and Freescale PowerPC. The DMA architecture based on FPGA is compatible with the Xilinx PCIe core while the DMA architecture based on POWERPC is compatible with VxBus of VxWorks. The solutions provide a high-performance and low-occupancy alternative to commercial. In order to maximize the PCIe throughput while minimizing the FPGA resources utilization, the DMA engine adopts a novel strategy where the DMA register list is stored both inside the FPGA during initialization phase and inside the central memory of the host CPU. The FPGA design package is complemented with simple register access to control the DMA engine by a VxWorks driver. The design is compatible with Xilinx FPGA Kintex Ultrascale Family, and operates with the Xilinx PCIe endpoint Generation 1 with lane configurations x8. A data throughput of more than 666 MBytes/s(memory write with data from FPGA to PowerPC) has been achieved with the single PCIe Gen1 x8 lanes endpoint of this design, PowerPC and FPGA can send memory write request to each other.
\end{abstract}


\section{Introduction}
For better observation,analysis and feedback, the data of scientific experiments such as Shanghai Synchrotron Radiation Facility(SSRF) and Shanghai Deep Ultra Violation Free Electric Laser(SDUV-FEL) need to be transmitted and processed online in real time \cite{beam_test}. Scientific systems contain data acquiring and processing subsystems, and the different subsystems will sustain Gigabyte per second data rates among them. Depending on the performance specifications of the particular application, FPGA with unique parallel processing and good timing control characteristics is usually used to acquire data from experiments, and PowerPC with powerful computational capacity is sometimes adopted as a host embedded system to process data on-line. However, the bottleneck between the subsystems usually lies in the data transmission link between FPGA and PowerPC. In our instrument, FPGA need to send data to PowerPC at the rate of no less than 500MBps. To solve such problem, we have developed PCI Express(PCIe) as data links.

PCIe is a widely used and reliable high speed data transmission protocol. Xilinx provides FPGA(XCKU040) with PCIe IP core(Gen3) \cite{Xilinx_PCIe_core}. Freescale provides PowerPC(MPC8641D) with PCIe integrated as a peripheral device \cite{PowerPC_PCIe_peripheral}, WindRiver supplies VxWorks which contains a Board Support Package(BSP) to compatible with PowerPC. Both FPGA and PowerPC supply users with basic PCIe device. To achieve the highest data throughput while reducing the occupancy of PowerPC, DMA engine is needed to overcome the limitations introduced by the PowerPC scheduling of operating system(OS). \cite{Xillybus_PCIe_DMA_core, PLDA_PCIe_DMA_core, Northwest_PCIe_DMA_core} implement DMA architectures for PCIe core based on FPGA of Xilinx, however they always cost too much and these architecture usually are limited to a single FPGA device. To conquer these restrictions, we designed and implemented a high-performance and compact DMA engine architecture which is fully compatible with Xilinx FPGA Ultrascale family, together with a custom-designed VxWorks driver based on VxBus. This paper illustrates the DMA engine architecture in FPGA, the VxWorks driver in PowerPC and the handshaking sequence between FPGA and PowerPC. The different payload length transmission are discussed. In this paper, two development kit boards are adopted: KCU105 is for XCKU040(FPGA), and HPCN8641D is for MPC8641D(PowerPC).

\section{System Overview of Point-to-Point PCIe}
\label{sec:system_overview}

PCIe is a layered protocol, containing a transaction layer, a data link layer, and a physical layer \cite{PCIE_layered_structure_wikipedia}. The structure of these layers for two different PCIe devices is illustrated in Figure \autoref{fig:PCIe_layered_structure}, \cite{PCIE_DMA_IEEE_Nuclear_Science} \cite{PCIE_layered_structure_wikipedia, PCIE_DMA_IEEE_Nuclear_Science}.

\begin{figure}[htb]
\includegraphics[width=3.5in,trim=75 30 45 20,clip]{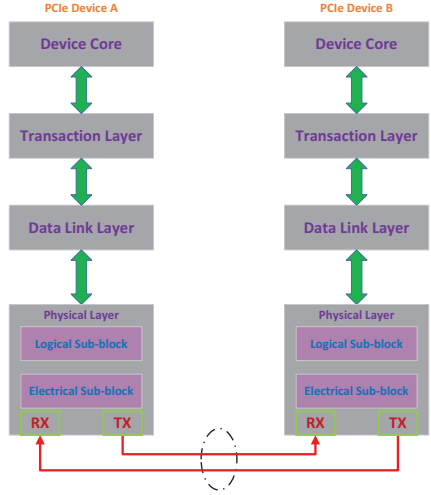}
\caption{PCIe Layer.}
\label{fig:PCIe_layered_structure.}
\end{figure}

\begin{figure}[htb]
\includegraphics[width=3.5in,trim=80 260 15 250,clip]{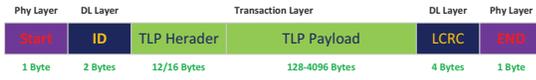}
\caption{PCIe Protocol.}
\label{fig:PCIe_protocol_structure}
\end{figure}

The Data Link(DL) Layer is subdivided to include a media access control (MAC) sub layer. The Physical Layer is subdivided into logical and electrical sub layers. The Physical logical sub layer contains a physical coding sub layer (PCS). The PCIe IP core of Xilinx comprises the DL layer and Physical layer and it supplies user with interface which complies with the bus of AXI-4 \cite{AXI_4} of transaction layer. It is convenient for users to consider only the transaction layer protocol(TLP) logic to finish PCIe transmission and Message Signaled Interrupts(MSI). The PCIe is a peripheral of the PowerPC and the BSP of the PowerPC contains a basic driver of PCIe, user will need to develop a compatible driver complied with VxBus for PCIe \cite{windriver_vxbus}.

Three kinds of memory spaces lie in PCIe: Memory Space, Configuration Space and I/O space. We mainly develop PCIe with memory space in this paper. The configuration space contains the PCIe configuration registers, each PCIe of device contains 6 base address registers(BAR 0-5), in this paper, only BAR0 is adopted and the memory size of BAR0 is configured as 2K bytes. The version of PCIe IP core  is Gen3 in XCKU040 while Gen1 in MPC8641D. In order to match the FPGA and PowerPC, considering the PCIe is downward compatible, the PCIe IP core in FPGA is configured as Gen1.

PCIe Gen1 offers a data link operating at 2.5 Gbps each lane and uses 8B/10B encoding, thus the actual maximum throughput of each lane is 2 Gbps. Additional packet overhead is also included excluding the basic payloads as the structure illustrated in Figure \autoref{fig:PCIe_packet_protocol_structure}.

We denote the maximum payload size with $P_{l}$, and the protocol header with $H_{pcie}$. $N$ is used as the number of lanes. The maximum theoretical throughput $V$ for PCIe Gen1 can be calculated with \autoref{equ:PCIe_Maximum_Speed} \cite{PCIE_DMA_IEEE_Nuclear_Science}.

\begin{equation}
\label{equ:PCIe_Maximum_Speed}
V = \frac{P_{l}}{P_{l}+H_{pcie}} \cdot N \cdot 250MBps
\end{equation}

\section{PCIe based on FPGA}
\label{sec:pcie_fpga}

\subsection{A High-Throughput DMA Architecture for PCIe Application}

\begin{figure}[htb]
\includegraphics[width=3.5in,trim=125 80 25 85,clip]{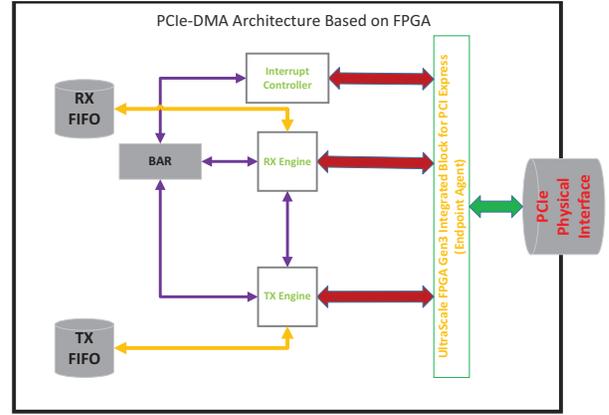}
\caption{Architecture of the DMA engine.}
\label{fig:FPGA_PCIe_DMA_Architecture}
\end{figure}

The DMA engine designed in this article adopts a stream mode in order to maximize the data throughput and minimize the FPGA resource utilization. The complete architecture of PCIe-DMA is shown in \autoref{fig:FPGA_PCIe_DMA_Architecture}.

FPGA is configured as the PCIe bus master to start memory write(MWR) and memory read(MRD) to PowerPC. The DMA uses TX engine to transmit data to host and RX engine to receive data from host. The PCIe core of XCKU040 offers an Advanced eXtensible Interface(AXI)-4. The width of AXI-4 interface is mainly configured as 128 bits. The AXI-4 contains four groups interface as following:

\begin{enumerate}[(1)]
\item   Completer Request Interface(CRI): User application receives completers from host via this interface group.
\item   Completer Completion Interface(CCI): User application replies the requester from host and delivers each TLP on this interface.
\item   Requester Request Interface(RRI): User application delivers each TLP of requester as an AXI4-stream packet.
\item   Requester Completion Interface(RCI): Host replies user requester and delivers the TLP to FPGA via this interface.
\end{enumerate}

User logic of TX engine is complied with RRI and CCI of PCIe Core in AXI4-Stream Slave mode, while the RX engine is complied with CRI and RCI. TX engine will send MWR and MRD TLP via RRI, then RX engine will receive requester completion information for MRD by RCI. Xilinx offers a basic PCIe communication example which implements MWR, MRD processing launched by host. We need to realize that MWR and MRD are launched by FPGA. The user clock in FPGA is configured as 125 MHz and the IP core is supplied with a reference clock 100 MHz from standard PCIe slot which is mounted on HPCN8641D.

\subsection{Base Address Register}

BAR0 in FPGA is implemented as a list of registers which are adopted as handshaking controlling. The width of each register is 32 bits according to the RISC width of PowerPC. The registers contains the following functions:

\begin{enumerate}[(1)]
\item   Initialization flag.
\item   MWR and MRD start flag.
\item   MWR and MRD interrupt processing flag: It indicates that the PCIe in FPGA generates a MSI and is waiting for the PowerPC finishing processing the interrupt.
\item   MWR address, payload length and MWR times.
\item   MRD address, payload length and MRD times.
\item   DMA MWR and MRD performance flag: It is used to indicate the current throughput of PCIe.
\end{enumerate}

\subsection{TX Engine}

A finite State Machine(FSM) diagram of the MWR DMA engine is shown in \autoref{fig:FSM_PCIe_MWR} and is described below, together with the handshaking sequence with the VxWorks driver:

\begin{figure}[htb]
\includegraphics[width=4.5in,trim=125 150 0 115,clip]{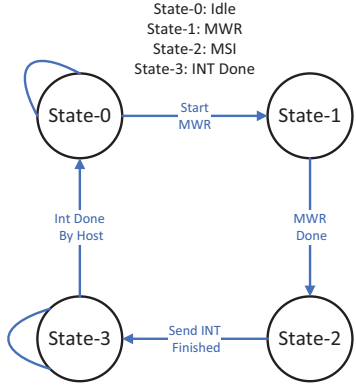}
\caption{FSM of  MWR DMA-PCIe in FPGA.}
\label{fig:FSM_PCIe_MWR}
\end{figure}

\begin{enumerate}[(1)]
\item   Initialization phase(state 0):PowerPC set the register values including times of Memory Write(MWR), payload length and initial address in BAR0 space.
\item   After the initialization phase, the FSM waits in idle(state 0) for Starting MWR(state 1).
    \begin{enumerate}
    \item   FPGA load the payload length and address to which it will send the data.
    \item   FPGA check the MWR times and update the WR pointer with the last loaded address.
    \end{enumerate}
\item   If the FPGA has already prepared the data for once DMA-TX, then the FPGA start moving data from FPGA to PowerPC(state 2).
\item   After finishing MWR, FPGA generate a MSI(state 2).
\item   If MSI is sent to IP core, then FPGA will wait for the int processed done signal from host(state 3).
\item   CPU deal with the MSI and return "interrupt done" to FPGA, then the FPGA return to idle(state 0).
\end{enumerate}

We denote the DMA MWR payload length(DWORDs number) with $PL_{mwr}$ and DMA MWR times with $N_{mwr}$. It will take one clock for FPGA to request once MWR, and the left clocks are valid data. The clock of user logic is 125MHz. Then the theoretical efficiency $F_{mwr}$  and speed $V_{mwr}$ of DMA is as \autoref{equ:FPGA_mwr_speed}.

\begin{subequations}
\label{equ:FPGA_mwr_speed}
\begin{align}
F_{mwr} &= \frac{PL_{mwr}}{PL_{mwr}+1},\label{equ:mwr_efficiency}\\
V_{mwr} &= \frac{PL_{mwr}*4}{(PL_{mwr}+1)*8}\cdot GBps \label{equ:mwr_speed}
\end{align}
\end{subequations}

Actually, we will count the real clock consumption as following: when $mwr_start_sig$ is valid, the counter will start to count, and when $mwr_done$ which indicates the termination of DMA-MWR is valid, the counter stops count. Each clock denotes 8ns, so it takes $8*counter_{value}$ ns. We denote the data size with $D_{mwr}$ bytes. The actual MWR speed will be calculated as \autoref{equ:real_mwr_speed}.

\begin{equation}
\label{equ:real_mwr_speed}
V_{mwr,real} = \frac{D_{mwr}}{8*counter_{value}}\cdot GBps
\end{equation}

\subsection{RX Engine}

A FSM diagram of the Memory Read(MRD) DMA engine is illustrated in \autoref{fig:FSM_PCIe_MRD} and is described below, together with the handshaking sequence with the VxWorks driver:

\begin{figure}[htb]
\includegraphics[width=4.5in,trim=125 150 0 100,clip]{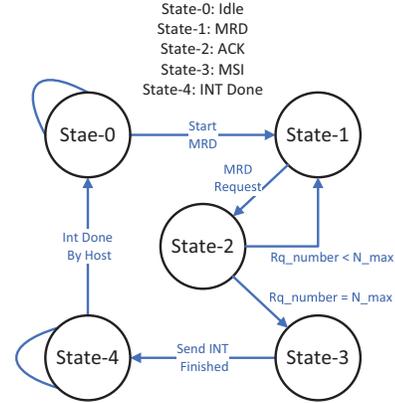}
\caption{FSM of  MRD DMA-PCIe in FPGA.}
\label{fig:FSM_PCIe_MRD}
\end{figure}

\begin{enumerate}[(1)]
\item   Initialization phase(state 0):PowerPC set the register values including times of MRD, payload length and initial address in BAR0 space.
\item   After the initialization phase, the PowerPC will prepare the data which will be moved from PowerPC to FPGA. Once the data has been ready, PowerPC will send a MRD command(state 0).
\item   If FPGA receives the MRD command, it will start the MRD(state 1).
    \begin{enumerate}
    \item   FPGA load the payload length and initial address.
    \item   FPGA send a MRD TLP to PowerPC.
    \end{enumerate}
\item   FPGA will wait for the ack from host, update the RD pointer with the last loaded address and check whether it is the last transmission of a MRD.
\item   After FPGA receives all the data from PowerPC. It will generate a MSI(state 3) and wait for the MRD stop command from VxWorks(state 2).
\item   If MSI is sent to IP core, then FPGA will wait for the int processed done signal from host(state 4).
\item   After FPGA receives the stop command, it returns to idle(state 0).
\end{enumerate}

We denote the DMA MRD payload length(DWORDs number) with $PL_{mrd}$ and DMA MRD times with $N_{mrd}$. It will take one clock for FPGA to request once MRD, and then the FPGA will receive the completion packet from PowerPC and the completion TLP contains one clock header for user logic to analyze and the left are valid payloads.  The clock of user logic is 125MHz.

Then the theoretical efficiency $F_{mrd}$  and speed $V_{mrd}$ of DMA is as \autoref{equ:FPGA_mrd_speed}.

\begin{subequations}
\label{equ:FPGA_mrd_speed}
\begin{align}
F_{mrd} &= \frac{PL_{mrd}}{PL_{mrd}+1},\label{equ:mrd_efficiency}\\
V_{mrd} &= \frac{PL_{mrd}*4}{(PL_{mrd}+1)*8}\cdot GBps \label{equ:mrd_speed}
\end{align}
\end{subequations}

Actually, we will count the real clock consumption as following: when the TLP in RX is valid, the counter will start to count, and when $mrd_done$ which indicates the termination of DMA-MRD is valid, the counter stops count. Each clock denotes 8ns, so it takes $8*counter_{value}$ ns. We denote the data size with $D_{mrd}$ bytes. The actual MRD speed will be calculated as \autoref{equ:real_mrd_speed}.

\begin{equation}
\label{equ:real_mrd_speed}
V_{mrd,real} = \frac{D_{mrd}}{8*counter_{value}}\cdot GBps
\end{equation}

\subsection{Interrupt Controller}

\begin{figure*}[htb]
\includegraphics[width=6in,trim=0 160 0 150,clip]{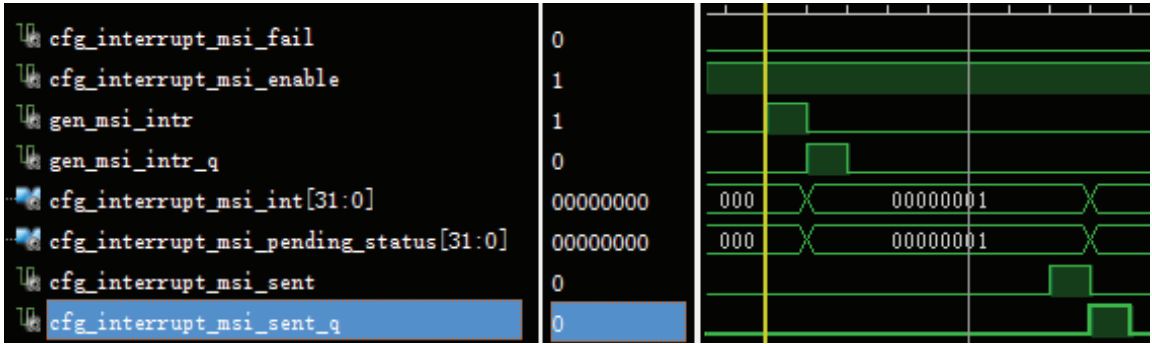}
\caption{MSI timing Diagram(AXI4-128) in FPGA.}
\label{fig:MSI_timing_digram}
\end{figure*}

It is convenient to generate an interrupt of MSI with Xilinx PCIe core by producing corresponding signals to the IP core, when the IP core feedback a sent signal, it means that the IP core has already successfully sent a MSI to PowerPC. Here, finishing both DMA-MWR and DMA-MRD will generate MSI, PowerPC will polling the MSI register which illustrates the current interrupt type in BAR0 space. The actual MSI timing diagram acquired by ILA is illustrated in \autoref{fig:MSI_timing_digram}, and it is the same as in \cite{Xilinx_PCIe_core}.

\section{PCIe based on PowerPC}
\label{sec:pcie_powerpc}

\begin{figure}[htb]
\includegraphics[width=3in,trim=0 80 0 60,clip]{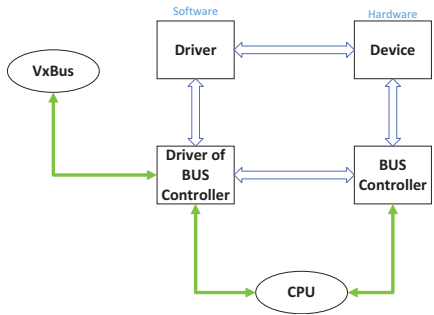}
\caption{Diagram of VxBus in VxWorks\cite{windriver_vxbus}.}
\label{fig:VxBus_Diagram}
\end{figure}

The BSP in VxWorks contains kernel, I/O system, file system and network support. WindRiver develop a brand new architecture which is called VxBus for driver development since VER 6.6. The relationship between VxBus and VxWorks is shown in \autoref{fig:VxBus_Diagram}.

VxWorks maintains two linked lists: driver list and devices list. After power is on, the initialization will match the drivers and devices, if they are matched, then the matched driver and device will combine to be an instance. The instance is also connected to an instance list which can be called by user application. Generally, user needs to register the custom-designed driver to VxBus via $vxbDevRegister()$ function. Then $pcieInstInit()$, $pcieInstInit2()$,and $pcieInstConnect()$ will be called to initialize PCIe device. After that, $pcieDMAInt()$ is defined as the interrupt service function and connected to the list of interrupt list.

The control registers that used as handshaking with FPGA is listed as head file in the driver. The registers is the same as registers in BAR0 of PCIe in FPGA. The buffers for TX and RX are both allocated for 2 Mbits. The driver also realizes MWR and MRD with payload length of double words(DWORD) to handshake with FPGA via PCIe.

The interrupt service routine(ISR) in this paper deals interrupt as following:

\begin{enumerate}[(1)]
\item   After initialization, the ISR will wait for coming of interrupt from MSI via PCIe.
\item   If the interrupt is generated, ISR will first read the interrupt status register in FPGA and check whether it is MWR or MRD interrupt.
\item   If it is DMA MWR done interrupt, PowerPC will move the data from the buffer in host to another address and process these data. After that, the PowerPC will send a MWR ISR done signal to FPGA. If the FPGA receives this ISR done signal, it will permit the next DMA MWR.
\item   If it is DMA MRD done interrupt, PowerPC will prepare new data to the MRD buffer for the next time transmission and send an ISR done signal to FPGA. If the FPGA receives this ISR done signal, it will permit the next DMA MRD.
\end{enumerate}

\section{Test Result}
\label{sec:result}

\begin{figure}[h]
\includegraphics[width=3in,trim=60 120 60 100,clip]{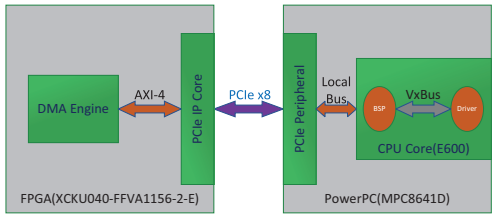}
\caption{System Overview.}
\label{fig:PCIe_System_Overview}
\end{figure}

The performance of the DMA engine has been measured using a Xilinx PCIe Gen1 core with different configurations. A KCU105 Kintex Ultrascale-PCIe board mounting on a HPCN8641D was used for the measurements with the Gen1 x8 lanes endpoints. The KCU105 contains a Xilinx XCKU040 device and HPCN8641D contains a Freescale MPC8641D device. The FPGA is plugged into the PCIe slot of the HPCN8641D. BSP is an original driver supplied by VxWorks and it is packaged as VxBus, user driver application is designed according to VxBus. The system architecture is shown in \autoref{fig:PCIe_System_Overview}.

The measurements don't take into account the initialization phase: the driver is loaded by the OS and registers are written into FPGA BAR0. The memory addressing is a 32-bit.

\begin{figure}[h]
\includegraphics[width=3in,trim=0 30 0 60,clip]{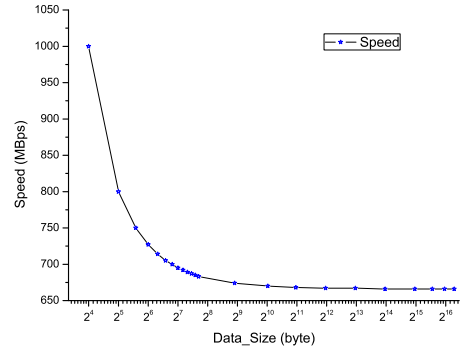}
\caption{MWR throughput vs. data size.}
\label{fig:mwr_with payload_16bytes}
\end{figure}

FPGA issue a MWR with a constant payload length of 16 bytes to PowerPC. There is one dummy clock beat between each MWR TLP. The measurements of this condition is issued in \autoref{fig:mwr_with payload_16bytes}.

\begin{figure}[h]
\includegraphics[width=3in,trim=0 45 0 60,clip]{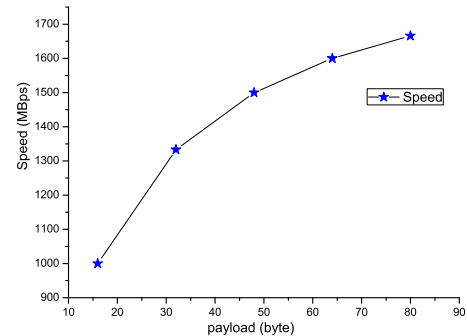}
\caption{Single MRD Completion TLP with different payload size.}
\label{fig:mrd_with_different_payload}
\end{figure}

We also issued a single MRD TLP from FPGA to PowerPC with different payload length. The result for receiving a single MRD completion TLP from PowerPC is shown in \autoref{fig:mrd_with_different_payload}.

The architecture of DMA-PCIe in this paper is lightweight. The resource requirements of this architecture are listed in \autoref{tab:FPGA_resource_requirements}.

\begin{table}[ht]
\centering
\caption{Resource Consumption on XCKU040. \label{tab:FPGA_resource_requirements}}
\begin{tabular}[c]{|c|c|c|c|} \hline
Resource & Estimation & Available & Utilization \\ \hline
LUT & 6689&242420 & 2.76\%   \\  \hline
LUTRAM  & 1174 &   112800  &   1.04\%  \\  \hline
FF & 11222   &   484800  &   2.31\% \\ \hline
BRAM & 34.5   &   600  &   5.75\% \\ \hline
\end{tabular}
\end{table}

\section{Conclusion}
\label{sec:conclusion}

A high-throughput PCIe with DMA engine based on FPGA and PowerPC was described in this paper, the DMA engine is compatible with the Xilinx Kintex Ultrascale PCIe Gen1 Core and a special PCIe driver complied with VxBus is implemented in VxWorks 6.6 based on Freescale MPC8641D. We can easily apply this work in real-time data acquiring and processing system.


The DMA architecture has implemented a high-throughput more than 666 MBps. Because there is a long time between the FPGA issue MRD and PowerPC answer the MRD TLP, we can not realize a more efficient DMA-MRD this time. Generally, this design is satisfied with our data transmission target. However, We didn't realize a MWR with payload more than 16 bytes this time, a more efficient DMA-MWR will be designed in future.

\end{document}